\documentclass[11pt,a4paper,fleqn]{article}

\begin{document}

\title{\bf Number of quantal resonances}

\author{Zafar Ahmed\footnote{zahmed@apsara.barc.ernet.in},
Sudhir R. Jain\footnote{srjain@apsara.barc.ernet.in}\\
Nuclear Physics Division,
Bhabha Atomic Research Center,\\
Trombay, Mumbai 400 085, INDIA}

\date{September 26, 2003}

\maketitle

\begin{abstract}
Based on extraction of resonances from  quantal time delay,
a theorem relating  quantal time-delay and the number of
resonances below a certain energy is proved here. Several illustrations
from quantum mechanics, neutron reflectometry and hadron resonances
are presented.
\end{abstract}


\vskip 0.5 truecm

Physics of weakly-bound systems and resonances has been of great interest for
the important role it plays in nuclear and particle physics.
In particular, deriving reliable information about  unstable and short-lived
states leads to a deeper understanding in the theory of elementary particles
in case of hadron resonances \cite{delta} and hypernuclei \cite{gal},
compound nuclear resonances \cite{blatt}
in $(n,\gamma)$ and $(p, \gamma)$ reactions and in development of new models 
and theories in the active field  of  radioactive nuclei \cite{tanihata}.
The research
on halo nuclei is also intimately connected with the search for superheavy
nuclei \cite{munzenberg}. The structure
of these elusive states is usually in the form of two-body, three-body or many-body
resonances \cite{nielsen}. It would be interesting if we could infer from 
the phase-shift data for $\pi$-$\pi$, or, $\eta$-$d$, or, $\alpha$-$\alpha$ 
scattering whether respectively a $\rho$-meson around $E=770 MeV$ is formed, 
or, a hyper-nucleus, or, a 
resonant $\alpha$-cluster \cite{wuos} is formed. Clearly, we would like to be sure of 
our analysis by studying simple, non-trivial illustrations from quantum 
mechanics. In this paper, we present various illustrations to elucidate 
how resonant structures can be reliably extracted and understood.    

The concept of central importance for our considerations is that of time-delay.
Time-delay, ${\cal T} $ is defined as the difference between the density of
states with and without interaction. This reduces to the following relation [5]
between ${\cal T} $ and the $S$-matrix \cite{goldberger},
\begin{equation} \label{1}
{\cal T} = -i\hbar \mbox{~tr~}S^{\dagger}\frac{dS}{dE}
\end{equation}
for a general form of interaction. For central potentials, for the $l$-th
partial wave, ${\cal T}_{l}$ reduces to $\hbar d\overline{\delta}_l/dE$ where
$\overline{\delta}_l$ is the difference between phase shift $\delta _{l}(E)$
and hard-sphere scattering phase-shift $\delta _{l}^{H}(a,E)$, and where $a$ is
the range of interaction. For an $s$-wave scattering,
$\delta _{l}^{H}(a,E) = -ka$  where $k$ is the wavenumber.
\par The complex-energy poles of time-delay on un-physical sheet like those of
$S$-matrix represent resonances. Consequently, the peaks
in ${\cal T}(E)$ are signatures of resonances \cite{brenig}.

In a system where number of single-particle states modify in the presence of 
an impurity, the change in the states is related to the energy-derivative of 
phase-shift. The total change in the density of states leads to the Friedel 
sum rule, which is basically a statement of charge-neutrality \cite{mahan}. 
On the other hand, an energy-integral of the phase-shift is shown to be 
connected to the energy of the impurity - Fumi's theorem. All the filled 
states contribute to the energy of the impurity via electron-impurity 
interactions. 

In the context of nuclear physics, some recent works \cite{alhassid} 
treat the effect of continuum by considering the contribution of naroow resonances. 
These narrow resonances are assumed to be like broadened bound states. This 
is, of course, not a valid assumption for broad resonances occurring in 
particle physics. When this assumption holds, the change in the single-particle 
continuum level density, $\delta \rho$, in the presence of a potential $V$ 
is found by subtracting the free particle level density $\rho _0$ from the 
total level density, $\rho$ :
\begin{equation}\label{2} 
\delta \rho = \rho (\epsilon )-\rho (\epsilon ) = \sum_{l,j} \frac{(2j+1)}{\pi}
\frac{d\delta _{lj}}{d\epsilon},
\end{equation}
the last being due to Friedel \cite{friedel}.  
The right hand side of (2) is connected with time-delay.   
We will consider an energy-integral of time-delay in this paper 
and illustrate it by a variety of physical situations of simple 
and advanced nature.   

For a narrow resonance at energy $E_0$ of width $\Gamma$, 
time-delay has a Lorentzian or a Breit-Wigner (B-W) form :
\begin{equation} \label{3}
{\cal T}(E)={{\hbar \Gamma/2} \over (E-E_0)^2+{\Gamma^2 \over 4}}.
\end{equation}
Then, it follows that the number of resonances below $E^\ast$ for a fixed partial wave 
is given by \cite{footnr}
\begin{eqnarray} \label{4}
n_R &=& {1 \over \pi}\int_{0}^{E^\ast}~ \left( {d\bar\delta(E) \over dE} \right)~dE \nonumber \\
&=&N+\Delta,~ 0< \Delta <1.
\end{eqnarray}
This is related to the approaches 
based on level-density \cite{alhassid,shlomo}. Since most of the methods fail to determine the
absolute values of phase-shifts,
therefore, R.H.S. of (4) would be different from the expression 
$[\bar\delta_l(E^\ast)-\bar\delta_l(0)]/\pi$. However, following somewhat more 
general considerations along the lines of \cite{tsang}, the spectral property 
of time-delay implies that  
\begin{equation} \label{5}
\int^{E^{\ast}} {\cal T}dE = n_R\hbar . 
\end{equation}
This is valid also for broad resonances, as shown in the Illustration IV below. 
To appreciate (5), following remarks are in order : (i) (5) is not the Levinson's  
theorem; (ii) if the energy-range includes bound states also, 
singularities will appear for bound states and non-zero widths appear 
for resonances, shapes being Lorentzians only for narrow ones 
(see also comments at the end of Illustration II below). 

Construction of time delay above using the complex energy-eigenvalues
may remind one of the potential scattering formalism due to
Kapur and Peierls \cite{kapur}. To illustrate, with the $s$-wavefunction in the range
of potential, the complex roots, ${\cal E}$ of
$\hbar \partial u/\partial r({\cal E},r=a) = i\sqrt{2m{\cal E}}u({\cal E},r)$ are taken as resonances
here. However, in \cite{kapur}, $\sqrt{2m {\cal E}}$ is replaced by
$\sqrt{2mE}$, where $E$ is fixed. Moreover, Kapur-Peierls' formalism
is designed to yield a non-unitary $S$-matrix.

At resonance, the phase-shift
$\bar \delta(E)=\arctan \left({\Gamma/2 \over E-E_0}\right)$, due to the
presence of an inverse trigonometric function, has  an essential ambiguity of $n\pi$
where $n$ remains uncertain. On the other hand, the ambiguity disappears
in ${\cal T}(E)$ as one gets a purely algebraic expression,
making  time-delay unambiguous and hence more reliable than the phase-shifts themselves.
The second derivative of $\bar\delta(E)$ reveals that on the either side
of $E=E_0$ the curvature changes sign and hence a point of inflexion in the variation
of $\bar\delta(E)$ is a necessary signature for the existence of a resonance.

Resonances which are  complex energy Gamow-Siegert
states of a potential generally arise  when
the potential is localized (e.g., short-ranged),or, when there is a barrier of a finite width
attached to the well. The square well represents the former when $l=0$ and the latter
when $l>0$. However, for simplicity we present s-wave square well
potential \cite{flugge}
and s-wave Dirac barrier potential $V_0\delta(r-a)$ \cite{flugge} which represents
the second situation. It is worth noting that the Illustrations III and IV 
discussed below are novel as they do not fall in the above categorization. Believing 
that a fundamental result is understood better in terms of diverse examples, we present them 
below. \\

\noindent
{\bf Illustration I }: {\bf Square-well potential} (for barrier, see \cite{foot3}) \\

To begin with, we consider the most familiar potential
$V(r<a)=-V_0,V(r\ge a)=0$,  as an example to illustrate our results
on resonances. It may be noted that a reliable description of narrow resonances 
is important for separating the background.  
The s-wave $S-$matrix can be written as
\begin{equation}\label{6}
S_0(E)={ik \tan pa+p \over ik \tan pa-p}.
\end{equation}
The complex poles (resonances) of $S_0(E)$, i.e., the roots of $ik \tan pa =p$,
are equivalent to demanding outgoing wave, $u_0(r \ge a)=e^{ikr}$, instead of
$u_0(r>a)$ as given above. The s-wave time delay for square-well potential
can be obtained as
\begin{equation}\label{7}
{\cal T}_0(E)=\hbar {V_0\tan pa+apk^2\sec^2 pa \over 2  pk(p^2+k^2\tan^2 pa)}.
\end{equation}
\par Relevant complex poles of S-matrix, $(E_j-i\Gamma_j/2)$, such that
$E_j, \Gamma_j>0$, are known to be the resonant energies. With the help of {\it FindRoot}
by {\it Mathematica}, by taking $2m=1=\hbar$, and fixing the values of $V_0$ and $a$,
we find the resonant energies (e.g., poles of (6), for $l=0$). We then construct
time-delay as a sum of Lorentzians as these resonances are narrow. 
The solid lines in Fig. 1 display the exact values of ${\cal T}_l(E)$
for $l=0,1,9,10$ for square-well
potential $(V_0=5, a=10)$ up to an energy, $E=10$. The dashed lines
show the time-delay as calculated by using first  fifteen complex energies
$E_j-i\Gamma_j/2$. Excellent reproduction of the time-delay
can be seen to testify the representation of the time-delay by a sum over 
Lorentzians 
for a localized potential possessing resonances. We check that $E_j$
are the energies where the peaks in time-delay occur. Further, we check that
$\Gamma_j$ is nothing but two times the inverse of the height of the $j^{th}$-peak.
The values of the integral of time-delay up to $E=10$, $n_R$ (4), are mentioned
in each part of Fig. 1 which can be verified to be giving the
number of resonances as
depicted by the number of peaks in ${\cal T}_l(E)$.
\par Interestingly, in the above calculations when $l=9,10$, we find that the
{\it first}
complex roots $E=0.38499-0.479894 i=E_A$ and $E=0.541725-0.574161 i=E_B$ respectively
do not show up  as peaks in ${\cal T}_l(E)$. Moreover, if they are included in the
summation (4), they spoil the reproduction of the time-delay at other energies.
Similarly, when $V_0=-5$ and $a=2$, we find that $E_1=0.023387-0.542466i=E_C$
and $E_2=9.38365-4.43007 i$
are the poles of $S_0(E)$ (6). We find that $E_1$ does not lead to
a peak in ${\cal T}_0(E)$, whereas $E_2$ shows up as a broad peak
in ${\cal T}_0(E).$ Notice that in these examples $\Gamma_j>E_j$ and the
corresponding B-W profile appears only partly for energies, $E>0$, most of
it creeps in the negative energy regime. We rule out $E_A,E_B,E_C$ as resonances
by studying the corresponding wavefunctions. The wavefunctions at
energies equal to the real part of these complex energies show 
scattering-state-like behaviour wherein the localization within
$r<a$ is inhibited.\\

\noindent
{\bf Illustration II :} {\bf Repulsive Dirac delta s-wave barrier}\\

Dirac delta barrier, $V_0 \delta (r-a),~V_0 > 0$ in three dimensional
Schr{\"o}dinger equation \cite{flugge}, even in s-wave, supports the 
Gamow-Siegert states. 
The poles of S-matrix can be worked out to be the complex roots of
$k\cot ka-aV_0=ik$,
a condition which one also gets by using the outgoing wave boundary condition
at $r=a$. The s-wave quantal time delay can be found as
\begin{equation}\label{8}
{\cal T}_0(E)=\hbar {aV_0\tan^2 ka+ak^2\sec^2 ka \over 2 [k^2\tan^2 ka+(k+aV_0\tan ka)^2]}.
\end{equation}
For $V_0=10$ and $a=1$, we find first four poles of S-matrix and construct
time-delay as a sum of Lorentzians (shown as dotted lines in Fig. 2). 
Even with only four poles, the ${\cal T}_0(E)$ (8)
(solid line) is excellently reproduced by the dotted line in Fig. 2. By an 
integration of time-delay
up to energy $E=170$, we get the value of the integral $n_R= 4.0114$, correctly
indicating four resonances or metastable states. In the limit of $V_0 \rightarrow \infty$,
we get $E_j \rightarrow {j^2\pi^2 \hbar^2 \over 2ma^2}$ and $\Gamma_j \rightarrow 0$
these states will be the bound states of a particle contained between two rigid walls
of width $a$. Thus, for large values of $V_0$ one gets Dirac delta like spikes
in ${\cal T}_0(E)$  at energies $E=E_j$.    \\

\noindent
{\bf Illustration III :}\\

While studying the reflectometry of the polarized neutrons from
magnetized superconductors, Zhang and Lynn \cite{zhang} arrived
at a novel result wherein  a single, pronounced, parameter-dependent dip
occurs in the reflectivity. In terms of simple quantum mechanics this means
that the reflectivity, $R(E)$,
of the potential-step barrier, $V(x)=V_1+V_2(1-e^{-x/a})$, as a function energy
shows a {\it single}, pronounced, spike-like minimum at an energy slightly above
the step, for  certain sets of parameters, $(V_1,V_2,a)$. Usually when a
semi-infinite potential is smooth, $R(E)$ is a smoothly converging function of
energy. A discontinuity of derivative at a point in a semi-infinite potential
turns out to be the essence \cite{ahmed1,ahmed2} of this novel reflectivity.
An interesting claim \cite{neutron} that the energy, $(E=E_d)$, at which this
reflectivity minimum occurs
corresponds to a `half-bound state' has been refuted by the help of
several model potentials \cite{ahmed2}. Let us take the case of an exponential step
where $r(E)=\sqrt{R(E)}e^{i\theta(E)}$ is given by
\begin{equation}\label{9}
r(E)= {ik J_{-2ipa}(2qa)+ q J^{\prime}_{-2ipa}(2qa) \over
ik J_{-2ipa}(2qa)-q J^{\prime}_{-2ipa}(2qa)},
\end{equation}
where $p={\sqrt{2m(E-V_1-V_2)} \over \hbar}$ and
$q=\sqrt{2mV_2 \over \hbar^2}$.
Whenever $2qa$ is close \cite{zhang} to the zeros of  $J_0(z)$, 
a single pronounced minimum exists in $R(E)$. More precisely, 
when $V_1=V_2=1$ and $a=1.31$ $(2m=1=\hbar)$ the dip in reflectivity occurs at
$E=2.0445=E_d$ \cite{ahmed1}. We find that reflection time-delay, ${d\theta \over dE}$,
possesses a single dip at, $E=E_d$, confirming the existence of a resonance
there (see Fig. 3). Since the contention is regarding the dip in reflectivity, 
we have calculated the reflection-phase-shift ($\theta$) and its derivative. 
Eventually, the integration (4) of time-delay over a long range of energy
(from  $E=2$ to E=4,6,10) divided by $\pi$ has been checked to yield a number 
very close to 1. Thus, the composite semi-infinite potential steps \cite{ahmed1,ahmed2}
are a new type of model potentials which are neither localized nor
there is a well attached to a barrier. They, however, possess a single
resonance. We also find that the low energy, single minimum in reflectivity
of the composite potential wells \cite{ahmed1} such as $V(x)=-V_0\exp(-|x|/a)$
is nothing but a resonance. \\

\noindent
{\bf Illustration IV :}\\

Usually, narrow resonances are treated as elementary
particles \cite{dashen}. However,
even relatively broader resonance like the excited state of a nucleon called
$\Delta ^{++} (1232)$ are treated as particles.  Here we would like to demonstrate that
time delay calculated from the available \cite{data} elastic scattering phase shifts
of $\pi ^{+}$ and $p$ ($P33$) yields a peak at $E=1218$ MeV of width $\Gamma=129$ MeV
(see Fig. 4) corresponding to the formation of the well-known
baryon resonance called the 
$\Delta$-resonance which has mass as 1232 MeV and width as 120 MeV. 
This result can also be found in \cite{kjk} where there is a  more
detailed study on unflavoured baryon resonances. 
The integration (4) of time-delay in the energy range given in Fig. 4
divided by $\pi$ yields $0.87$ instead of unity,
simply because the resonance data is truncated at the lower energies.
This illustration, in addition to \cite{kjk} shows that (4) and (5) 
are also valid for non-central interactions and for broad resonances 
with lifetimes $\sim 10^{-23}$s. It should also be noted that one is 
in a relativistic regime with energies of the order of GeV. The concept 
of time-delay holds in relativistic regime because of its connection with 
the $S$-matrix (1) \cite{dmb}. 

Hitherto, the
energy-integral of the first derivative of phase shift over a long range of energy
is supposed to either vanish \cite{wigner} or yield the number of bound states
\cite{dashen} via the well-known result in (10) below. 
In such a confusing scenario, we have
brought  out the correct meaning of the energy-integral mentioned above.
We have revealed that  the integral (4)  does not vanish and it  yields the  number
of resonances possessed by a potential. Since time and energy are conjugate variables,  
(5) can be interpreted as  an analogue of the Bohr-Sommerfeld
quantization. This is also a quantum analogue of
some results  in classical scattering theory obtained twenty
years ago by Narnhofer and Thirring \cite{narnhofer}. 

Finally, we would like to make some observations on the Levinson's theorem 
in its original form  relates the phase-shift at zero energy to the number 
of bound states, 
$n_B$, possessed by an attractive
potential : $n_B=\delta_l(0)/ \pi $ \cite{goldberger,foot1}.
However, it  is very
often written as \cite{goldberger,dashen,foot2,bolle}
\begin{equation} \label{10}
n_B=[\delta_l(0)-\delta_l(\infty)]/\pi .
\end{equation}
It is important to note that (10) seems to relate the 
negative-energy bound
states to the phase-shift at an infinitely positive energy. This is basically 
done to provide a reference at infinite energy  where phase shift is 
assumed to be zero.  
Although for short-range potentials, $\delta_l(\infty)$ is zero, this 
is not generally true \cite{foot2}. In a specific calculation, one may 
find $\delta_l(0)$ and $\delta_l(\infty)$ with ambiguous factors of, 
say, $m_1\pi$ and $m_2\pi$ with $m_1, m_2$ as two arbitrary, uncontrollable 
integers, having nothing to do with the potential. To obtain phase shifts 
free of such ambiguities (modulo $\pi$), one needs to employ special methods 
like the variable phase approach \cite{calogero} and an integral 
representation of phase shift found recently \cite{chadan}.

\newpage
\noindent
{\bf Figure Captions :} \\
Fig. 1(a): Time-delay, ${\cal T}(E)$, for a square well-potential
$(V_0=-5,a=10)$ for four values of angular momentum $l=0,1,9,10$.
The solid lines are the exact calculation and the dashed lines represent 
a sum of Lorentzians  using first fifteen resonances (poles of
$S$-matrix on un-physical sheet). Interestingly, in the case of $l=9,10$
the very first complex energies $E_A=0.38499- 0.47984 i$ and
$E_B=0.541725-0.574161 i$, respectively are not included
(see text for details). Notice that the values of $n_R$ agree with the
number of peaks in all the plots of time-delay, illustrating (4).

Fig. 1(b): The first peaks corresponding to Fig. 1(a) are separated 
out and shown for clarity.  

Fig. 2: Excellent reproduction of the time-delay (8) by a sum of 
Lorentzians for 
the Dirac delta potential
by using just first four  complex energies (resonances). A value of $n_R$
close to 4, illustrates existence of four resonances up to $E=170$.

Fig. 3: An instance of reflectivity, $R(E)$, dip in the exponential semi-infinite potential
step when $V_1=V_2=1$ and $a=1.31$. At $E=2.0445$ there occurs a dip in R(E) and
${\cal T}(E)$  and the 
(reflection) phase-shift, $\theta(E)$ displays point of inflexion.
The value of $n_R$ from $E=2$ up to $E=4,6,10$ has been found to be very close
to 1.

Fig. 4: The well known baryon resonance, $\Delta^{++}$ \cite{delta} is demonstrated
as a peak in time-delay in the phase-shift data \cite{data} of the scattering :
$\pi^++p \rightarrow \Delta^{++}$. We get $M_{\Delta}=1218$ MeV and
$\Gamma_{\Delta}=129$ MeV, whereas the standard values are $M_{\Delta}=1232$ MeV
and $\Gamma_{\Delta}=120$ MeV. The value of $n_R$ we get is $0.87$.


\begin{thebibliography}{99}
\bibitem{delta} T. Ericson and W. Weise, {\it Pions and Nuclei}
(Clarendon Press, Oxford, 1988).
\bibitem{gal} A. Gal, Adv. Nucl. Phys., Eds M. Baranger and E. Vogt,
vol. 8, 1 (1975) (Plenum Press, London); R. H. Dalitz et al. Proc. Roy.
Soc. Lond. A{\bf 426}, 1 (1989).
\bibitem{blatt} J. M. Blatt and V. F. Weisskopf, {\it Theoretical Nuclear
Physics} (John Wiley $\&$ Sons, London, 1952).
\bibitem{tanihata} I. Tanihata et al., Phys. Rev. Lett. {\bf 55}, 2676
(1985); A. Ozawa, T. Suzuki, I. Tanihata, Nucl. Phys. A{\bf 693},
32 (2001).
\bibitem{munzenberg} G. Munzenberg, Il. Nuovo Cim. {\bf 111A}, 747 (1998).
\bibitem{nielsen} E. Nielsen, D. V. Fedorov, A. S. Jensen, E. Garrido,
Phys. Rep. {\bf 347}, 373 (2001).
\bibitem{wuos} A. H. Wuosmaa et al., Phys. Rev. Lett.  {\bf 68}, 1295 (1992);
E. T. Mirgule et al., Nucl. Phys. A {\bf 583}, 287 (1995);
W. D. M. Rae and A. C. Merchant, Phys. Rev. Lett. {\bf 74}, 4145 (1995).
\bibitem{goldberger} M. L. Goldberger and K. M. Watson,
{\it Collision Theory}  (John-Wiley and Sons, Inc., New York, 1964).
\bibitem{brenig} W. Brenig and R. Haag,
Fortschritte der Physik {\bf 7}, No. 4/5, 183 (1959); also in
{\it Quantum Scattering Theory} edited by M. Ross  (Indiana
University Press, Bloomington, 1963) pp. 1-108.
\bibitem{mahan} G. D. Mahan, {\em Many-Particle Physics} (Plenum Press, New York, 1981). 
\bibitem{alhassid} Y. Alhassid, G. F. Bertsch, and L. Fang, preprint 
 available from {\sf www.lanl.gov} as eprint nucl-th/0303040; 
K. Bennaceur, J. Dobaczewski, M. Ploszajczak,
Phys. Rev. C{\bf 60}, 034308 (1999). 
\bibitem{friedel} J. Friedel, Proc. Roy. Soc. London {\bf 43}, 153 (1952). 
\bibitem{footnr} For resonances with $\Gamma \ll E_0$, the $S$-matrix is \cite{baz}\\ 
$S(E)=\exp[2i\bar \delta (E)]=\left ( {E-E_0+i\Gamma/2 \over E-E_0-i\Gamma/2}\right)$. 
\bibitem{shlomo} S. Shlomo, V. M. Kolomietz, and H. Dejbakhsh, 
Phys. Rev. C{\bf 55}, 1972 (1997). 
\bibitem{tsang} T. Y. Tsang and T. A. Osborn, 
Nucl. Phys. A{\bf 247}, 43 (1975). 
\bibitem{kapur} P. L. Kapur and R. E. Peierls, Proc.Roy. Soc., A166 (1938) 277.
\bibitem{flugge} S. Fl\"{u}gge, {\it Practical Quantum Mechanics}
(Springer-Verlag, Heidelberg, 1970).
\bibitem{foot3} For a square barrier, we find resonances above the top of the barrier 
using time-delay and verify our result for the number of resonances. 
\bibitem{zhang} H. Zhang and J.W. Lynn, Phys. Rev. Lett. {\bf 70} (1993) 77;
Phys. Rev. B {\bf 38} 15893 (1993).
\bibitem{ahmed1} Z. Ahmed, Phys. Lett. A {\bf 210} (1996) 1;
J. Phys. Math. Gen. A {\bf 32} (1999) 2767;
Phys. Lett. A {\bf 236} (1997) 289.
\bibitem{ahmed2} Z. Ahmed, J. Phys. Math. Gen. {\bf A33} 3161 (2000).
\bibitem{neutron} M. Geoghegan and G. Jannink,
Proc. Roy. Soc. Lond.  {\bf A454} 659 (1998).
\bibitem{dashen} R. F. Dashen and R. Rajaraman, Phys. Rev. {\bf D10},
694; 708 (1974).
\bibitem{data} B. H. Bransden, D. Evans, J. V. Major, {\it The Fundamental Particles}
(Van Nostrand Reinhold Company, London, 1973).
\bibitem{kjk} N. G. Kelkar, M. Nowakowski,  K. P. Khemchandani, S. R. Jain,
``Time-delay plots for unflavoured baryons'', preprint available from 
{\sf www.lanl.gov} as eprint hep-ph/0208197.
\bibitem{dmb} R. Dashen, S. K. Ma, and H. J. Bernstein, Phys. Rev. 
{\bf 187}, 345 (1969). 
\bibitem{wigner} E. P. Wigner, Phys. Rev. {\bf 98} (1955) 145.
\bibitem{narnhofer} H. Narnhofer, Phys. Rev. D{\bf 22}, 2387 (1980); 
H. Narnhofer and W. Thirring, Phys. Rev. A{\bf 23}, 1688 (1981).
\bibitem{foot1} For the case  when the edge of the potential separating
bound and scattering states ($E=0$) possesses a ``half-boundstate'', one
writes $n_B=\delta_l(0)/ \pi - 1/2$.
\bibitem{foot2} $\delta{(\infty)}$ does not vanish for several classes of
potentials (see e.g. Appendix III in \cite{calogero}).
\bibitem{bolle} D. Boll\'{e} and T. A. Osborn, J. Math. Phys.
{\bf 20}, 1121 (1979). 
\bibitem{calogero}  F. Calogero, {\it Variable Phase Approach to
Potential Scattering} (Academic, New York, 1967).
\bibitem{chadan} K. Chadan, R. Kobayashi, and T. Kobayashi, 
J. Math. Phys. {\bf 42}, 4031 (2001).
\bibitem{baz} A. I. Baz, Ya. B. Zel'dovich and A. M. Perelomov, 
{\it Scattering,
Reaction and Decay in  Non-relativistic Quantum Mechanics} (IPST Press, Jerusalem, 1969).
\end{thebibliography}
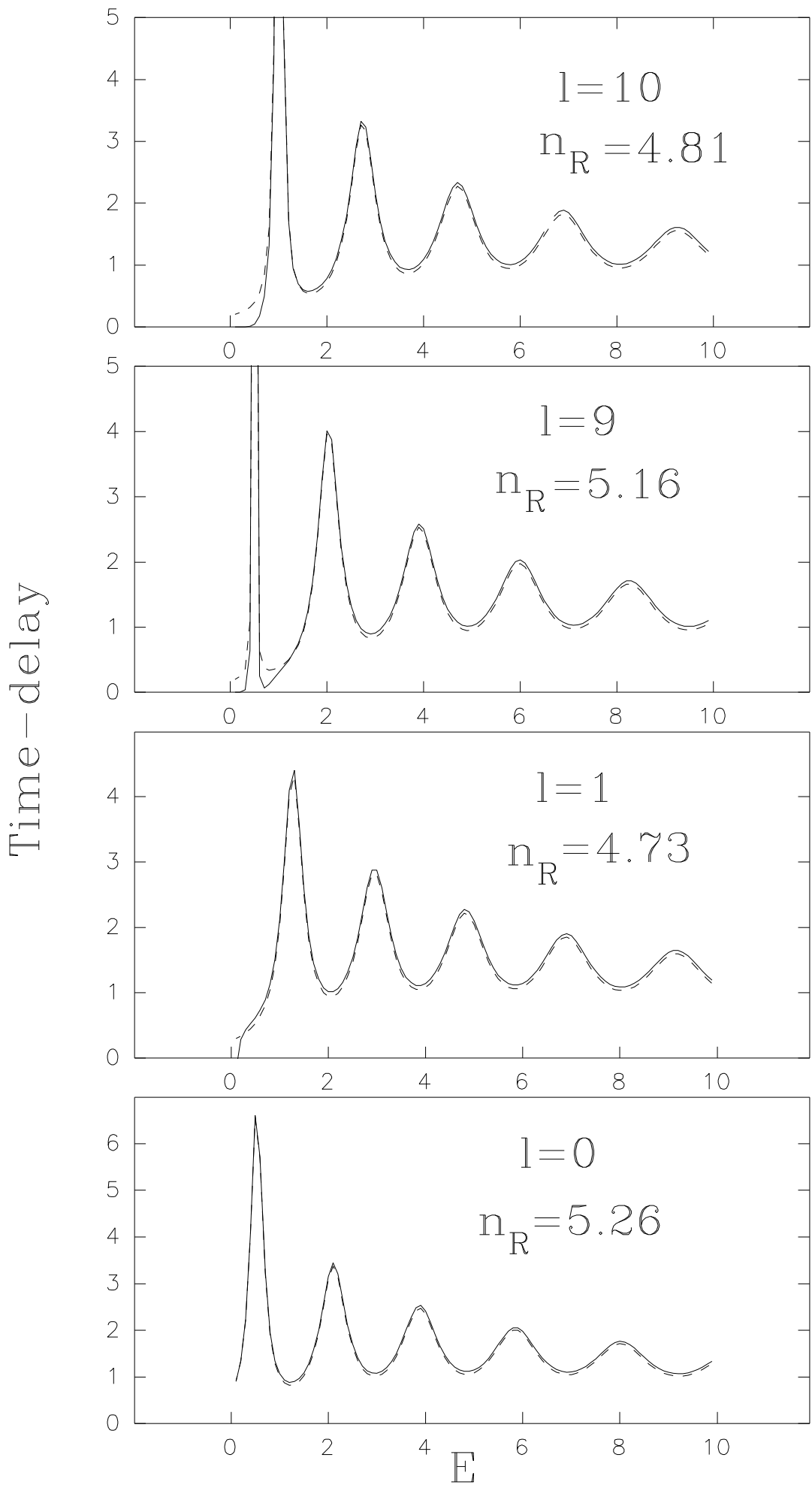
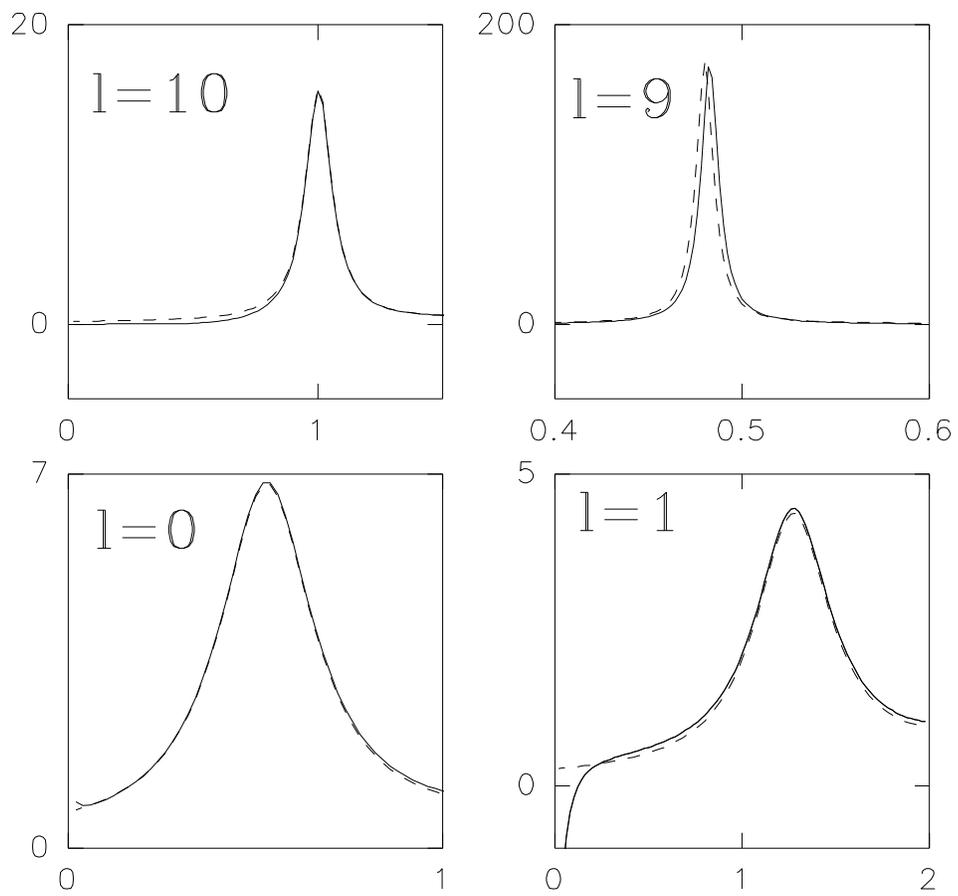
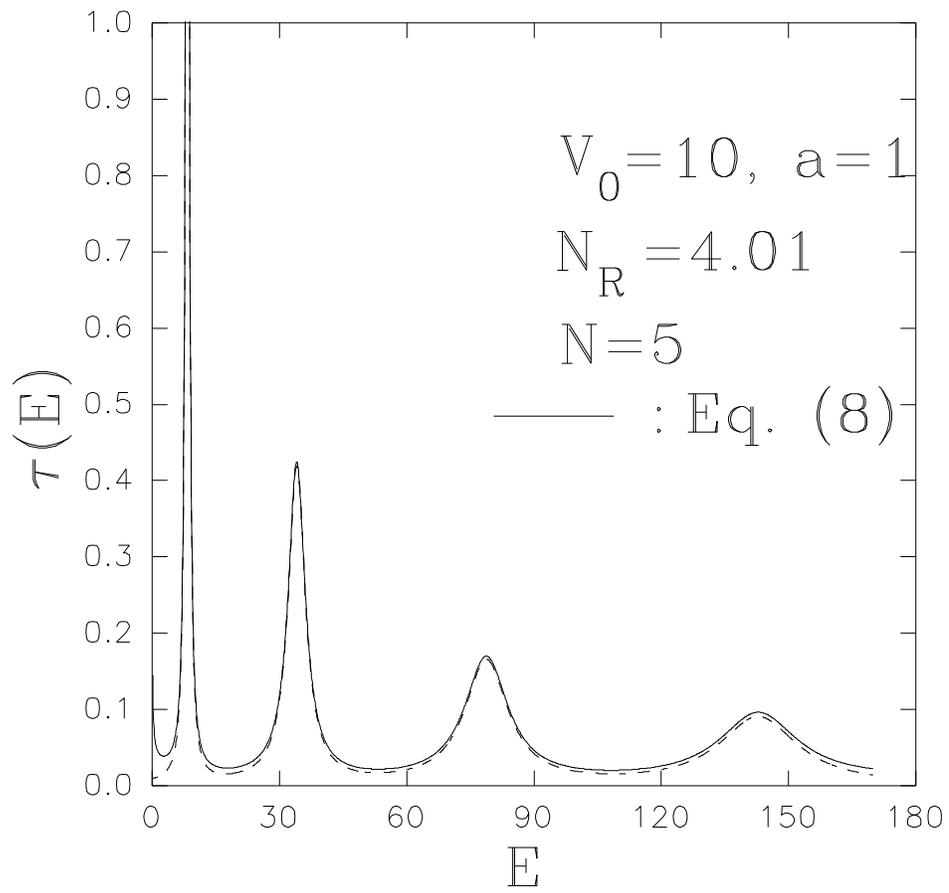
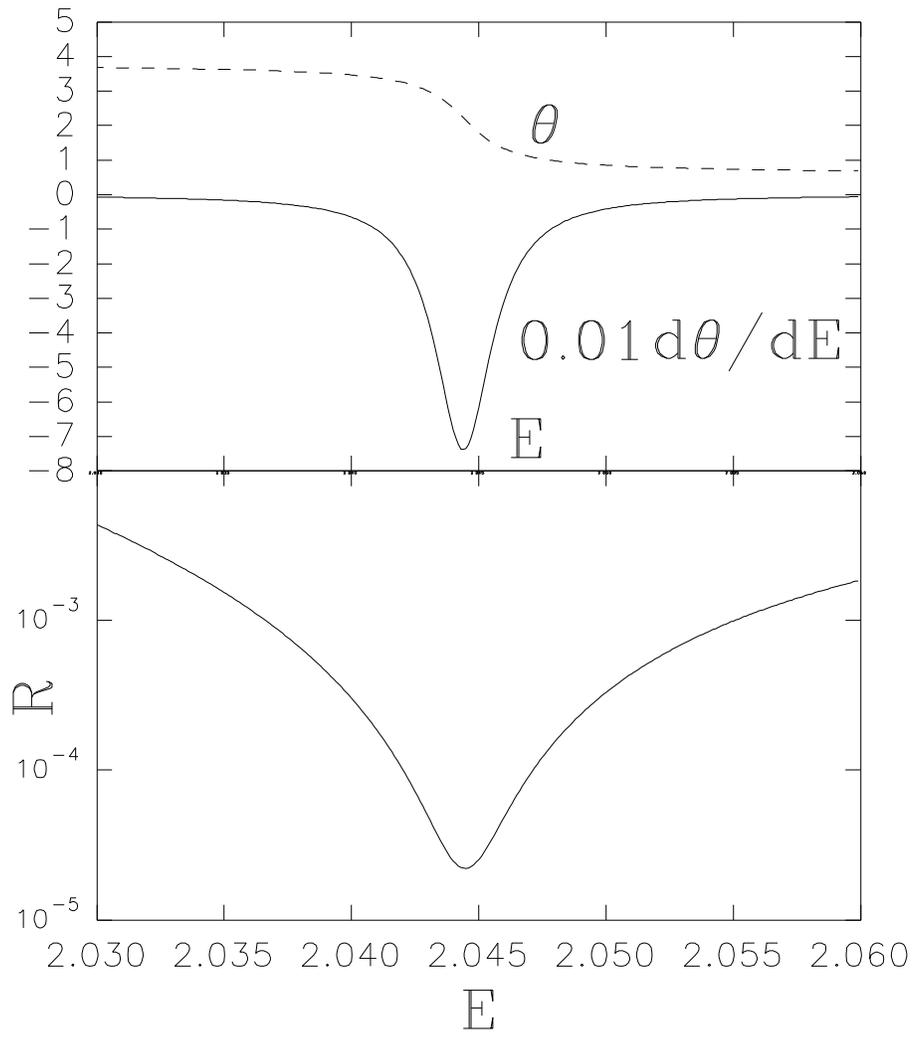
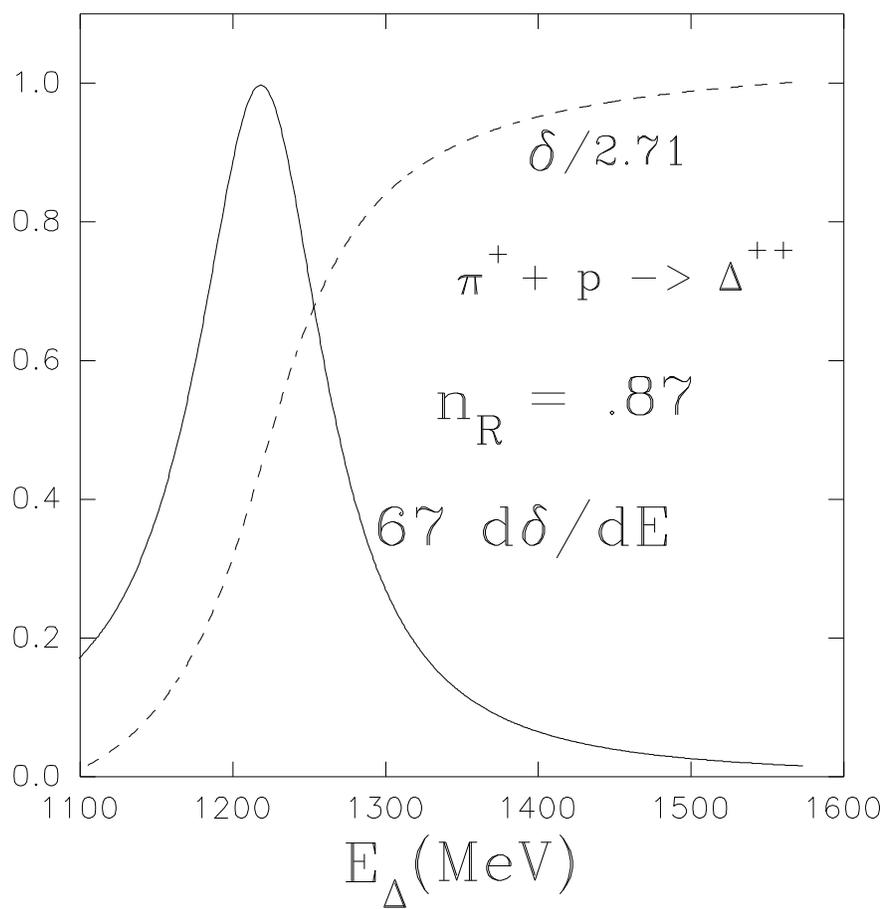
\end{document}